\documentclass[aps,prd,twocolumn,nofootinbib]{revtex4-1}

\usepackage{amssymb}
\usepackage{amsmath}
\usepackage{graphicx}
\usepackage[margin=5pt, font=small,labelfont=bf,justification=raggedright]{caption}

\usepackage{url}
\usepackage[bookmarks, pagebackref=false]{hyperref}
\usepackage{color}
	\definecolor{rossoCP3}{cmyk}{0,.88,.77,.40}
		\definecolor{graa}{rgb}{0.8,0.8,0.8}
		\definecolor{blaa}{rgb}{0.2,0.2,0.6}
		\definecolor{gron}{RGB}{0,150,0}
		\hypersetup{
			colorlinks, 
			bookmarksopen, 
			bookmarksnumbered,
			citecolor=blaa, 		
			linkcolor=black, 
			urlcolor=blaa,			
			}

\newcommand{\ea}[1]{%
\begin{align}
#1
\end{align}
}

\reversemarginpar

\def\Imt{{\rm Im}\tau}
\def\Re{{\rm Re}}

\def\G{\mathcal{G}}

\def\Ord{\mathcal{O}}

\begin{document}

\null\hfill\hbox{NORDITA-2019-071}\\[-2mm]

\title{\Large 
On the multiloop soft theorem of the dilaton in the bosonic string}

\author{
Paolo Di Vecchia$^{a,b}$,
Raffaele Marotta$^{c}$,
Matin Mojaza$^{d}$
}

\affiliation{
\vskip .3cm
$^a$ 
 NORDITA, KTH Royal Institute of Technology and Stockholm 
University\\
Roslagstullsbacken 23, SE-10691 Stockholm, Sweden \\
$^b$ The Niels Bohr Institute, University of Copenhagen, 
Blegdamsvej 17, DK-2100 Copenhagen \O , Denmark\\
 $^c$  Istituto Nazionale di Fisica Nucleare (INFN), Sezione di Napoli, 
Complesso Universitario di Monte S. Angelo ed. 6, via Cintia, 80126, Napoli, Italy\\
$^d$ Albert-Einstein-Institute, Max-Planck-Institute for Gravitational Physics, 
Am M\"uhlenberg 1, DE-14476 Potsdam-Golm, Germany
}

\begin{abstract}
In this note we show that by fixing the multiloop Green function in the closed bosonic string to be Arakelov's Green function, one obtains factorization of scattering amplitudes with a softly emitted dilaton to the same level as with a graviton to all loop order.
This extends our previous analysis at one loop to all loop orders and confirms that some high-energy quantum symmetry in the bosonic string protects the factorization of amplitudes with softly emitted dilatons.
\end{abstract}

\maketitle

\section{Introduction}\vspace{-3mm}

In \cite{DiVecchia:2018dob} we showed that in the bosonic string
the $h$-loop amplitude of $n$ closed tachyons, carrying momentum $k_i$, and one massless closed string with polarization $\epsilon^\mu_q  {\bar{\epsilon}}^\nu_q$, carrying momentum $q$, is given by:
\ea{
 &M_{n, 1}^{(h)} = C_h N_0^{n} \int d\mu_h^{(d)} \int_{\Sigma_h} \prod_{i=1}^n d^2 z_i   \prod_{i<j=1}^n  {\rm e}^{ \frac{\alpha'}{2} k_i k_j  \G_h (z_i , z_j ) }
 \nonumber \\
& \ \ \times
N_0 \epsilon^\mu_q  {\bar{\epsilon}}^\nu_q \int_{\Sigma_h} d^2 z \prod_{\ell=1}^n
{\rm e}^{ \frac{\alpha'}{2}  k_{\ell} q \G_h (z, z_{\ell})} 
I_{\mu \nu} (z_i, z; k_i, q) \, ,
\label{Mn1h}
}
where $C_h$ and $N_0$ are normalization factors, $d\mu_h^{(d)}$ is the integration measure of the moduli space of dimension 1 for $h=1$ and dimension $3h-3$ for $h\geq2$ including a factor arising from compactifying $26-d$ spatial dimensions. 
Their explicit form, irrelevant for the current discussion, were derived in~\cite{DiVecchia:2018dob} using Schottky parametrization.
The Koba-Nielsen variables $z_i$ and $z$ are integrated over the genus $h$ compact Riemann surface, $\Sigma_h$.
$\G_h$ is the two-point Green function, 
considered in more detail shortly.
Finally, the integrand $I_{\mu \nu}$ is given by:
\ea{
I^{\mu \nu} &= I^{\mu \nu}_{h} + 2 \pi \, h\, \eta^{\mu \nu} \mathcal{K}_h
\, ,  \label{2}\\
I^{\mu \nu}_{h} &=  \frac{\alpha'}{2} \sum_{i,j=1}^n k_{i}^{\mu} k_{j}^{\nu}   \partial_z \G_h (z, z_i ) \partial_{\bar{z}} \G_h (z, z_j )
\, , \\
\mathcal{K}_h& = \frac{1}{4 \pi h} \sum_{I,J=1}^h  \omega_I (z) ( 2\pi \Imt )^{-1}_{IJ}{\bar{\omega}}_J (z) \, ,
}
where $\omega_I$ and $\tau_{IJ}$ are the abelian one-forms and the period matrix of $\Sigma_h$, whose explicit expressions in terms of the Schottky group parameters were also given in \cite{DiVecchia:2018dob}.
Eq.~\eqref{Mn1h} is also valid at genus $h=0$ with $\mathcal{K}_0 = 0$ (and $\G_0(z,w) = \ln |z-w|^2$) and with the measure properly replaced by one over the $SL(2, \mathbb{C})$ volume-form, cf.~\cite{DiVecchia:2015oba}.

As explicitly shown in~\cite{DiVecchia:2018dob}, $\mathcal{K}_h$ has the property that
\ea{
\int_{\Sigma_h} d^2 z \, \mathcal{K}_h = 1\, ,
\label{intK}
}
showing that $\mathcal{K}_h\, d^2 z $ forms a genus $h$ unit volume form.
In fact, $\mathcal{K}_h$ is the metric induced from the pullback of the K\"ahler form from the Jacobian variety of $\Sigma_h$, a K\"ahler flat manifold,
as recently discussed in~\cite{DHoker:2017pvk,Basu:2018bde} in relation to modular graph functions.
We will here make important use of this fact.

 In \cite{DiVecchia:2018dob} we showed that when restricted to symmetric polarizations of the massless closed string the $z$-integral in \eqref{Mn1h}
 is by a soft expansion in $q$ fully localized through the order $q$, apart from the $\eta^{\mu \nu}$-term at order $q$, by only using generic properties of the multiloop Green function, in particular its Laplacian
 \ea{
 \partial_z \partial_{\bar{z}} \G_h(z,w) = \pi \delta^{(2)}(z-w) + r_h(z) \, ,
 }
 where $r_h(z)$ is a genus-dependent function ensuring Gauss' law on $\Sigma_h$, i.e. $\int d^2 z  \partial_z \partial_{\bar{z}} \G_h(z,w) = 0$.
 More explicitly, we were able to show that,   on the support of momentum conservation, for any $h$:
 \ea{
&\int_{\Sigma_h} d^2 z \prod_{\ell=1}^n
{\rm e}^{ \frac{\alpha'}{2}  k_{\ell} q \G_h (z, z_{\ell})} 
\, I^{\mu \nu}_{h} (z_i, z; k_i, q) 
\nonumber \\
&= f^{\mu \nu} (\G_h(z_i, z_j); k_i, q) + \Ord(q^2) \, ,
 }
where $f^{\mu \nu}$ is a universal function whose $z_i$ and genus dependence  enter only implicitly through the Green function in its argument.
This is a remarkable result reflecting a large number of nontrivial cancellations taking place among all 
terms that cannot be localized on the $\delta$-function by repeated use of integrations by part.
These cancellations are an imprint of some underlying symmetry of the theory.
They are, however, not totally surprising; 
the soft graviton theorem generates, in fact, $f^{\mu \nu}$, i.e.
\begin{widetext}
\ea{
 \prod_{i \neq j }^n
{\rm e}^{ \alpha'  k_{i} \cdot k_j  x }  \ f^{\mu \nu} ( x ; k_i, q) = 2 \pi \sum_{i=1}^n \left [\frac{k_i^\mu k_i^\nu}{k_i \cdot q}
- i \frac{q_\rho k_i^\mu L_i^{\nu \rho}}{k_i \cdot q}
- \frac{q_\rho q_\sigma }{2 k_i \cdot q}: L_i^{\mu \rho} L_i^{\nu \sigma} :
\right ]  \prod_{i \neq j }^n
{\rm e}^{ \alpha'  k_{i} \cdot k_j  x } \, ,
}
\end{widetext}
where $L_i^{\mu \nu} = i (k_i^\mu \partial_{k_i}^\nu - k_i^\nu \partial_{k_i}^\mu)$ is the (orbital) angular momentum operator, and the notation $: \ :$ means that derivatives are normal ordered (all act to the right).
It is thus an explicit demonstration of the recently extended soft theorem of the graviton~\cite{1404.4091, BDDN, BDPR,Laddha:2017ygw}
 at the multiloop level in the  bosonic string.

This result, however, goes beyond the graviton soft theorem, since  tracelessness in the polarization of the external massless closed string state was not assumed. This is also why normal ordering is important in the subsubleading generator of $f^{\mu \nu}$.

At genus zero, this is all there is, thus establishing a unified soft theorem for both the graviton and dilaton at the tree level~\cite{DiVecchia:2015oba,DiVecchia:2016amo}. But at higher genus the additional term proportional to $\eta^{\mu \nu}$ in $I^{\mu \nu}$ becomes relevant for the soft behavior of the dilaton, and this is the term we here wish to discuss. 
We will do this by first reviewing the genus one case in a slightly different way than outlined in~\cite{DiVecchia:2018dob}. Subsequently, we will generalize the analysis to all-loop order, by making use of the Arakelov construction for the Green function~\cite{DHoker:1989ima,Wentworth}.

\section{Soft factorization at genus one}

The genus one Green function in the Schottky parametrization
 reads
(see Appendix~\ref{SJ} for derivations and expressions) 
\ea{
\G_1(z_1,z_2) =
&\log \Big | \frac{ \theta_1( z_1/z_2 | \kappa ) }{\theta_1'(0 |  \kappa) } \Big |^2
+ \log \Big | \frac{(2\pi  z_1)(2 \pi  z_2) }{V_1'(0)V_2'(0)} \Big |
\nonumber \\
&
-\frac{1}{2 \pi \tau_2} \log^2 \Big | \frac{z_1}{z_2} \Big | \, ,
\label{G1}
}
where $\theta_1$ is the Jacobi theta function (eq.~\eqref{thetaprime})
and $\kappa$ is the Schottky group multiplier. 
The modular parameter on the torus, $\tau = \tau_1 + i\tau_2$, here arises as the `dimension one' period matrix and is related to the multiplier by the identity: $\kappa = e^{2\pi i \tau}$.
$V_i(z)$ are projective transformations
that define the local coordinates around each puncture $z_i$ on the Riemann surface (here the torus) such that $V_i (0)=z_i$ and inversely $V_i^{-1}(z_i) = 0$. The second term thus expresses
the `gauge freedom' due to worldsheet diffeomorphism invariance, and by inspection of \eqref{Mn1h} it is evident that the amplitude is independent of this term on shell and on support of momentum conservation.

There is a coordinate choice which gets rid of the second term above altogether, reading $V_i(z) = 2 \pi z_i \, z + z_i$, with its inverse being $V_i^{-1} = (z-z_i)/(2\pi z_i)$, thus satisfying the coordinate conditions. Since then $V_i'(0) = 2\pi z_i$, it follows that the second term vanishes, yielding the usual bosonic Green function on the torus, which is translationally invariant in the variables $\nu_i$, defined through $z_i= e^{2 \pi i \nu_i}$, i.e.
\ea{
G_B(\nu_{12} | \tau) =  
\log \Big |  \frac{ \theta_1( \nu_{12} | \tau ) }{\theta_1'(0 |  \tau) } \Big |^2
-\frac{2\pi}{\tau_2} (\text{Im}(\nu_{12}))^2 \, ,
}
where $\nu_{12} = \nu_1 - \nu_2$.
There is, however, an even more useful choice of coordinates, which reduces the Green function to a certain lattice sum, namely by choosing $V_i$ such that 
$V_i'(0) = z_i/(2\pi |\eta(\tau)|^2)$, where \mbox{$\eta(\tau) = [\theta_1'(0 |  \tau)/(2\pi)]^{1/3}$} is the Dedekind eta function,
whereby the Green function instead takes the form:
\ea{
G_A(\nu_{12} | \tau) =  
\log \Big | \frac{ \theta_1( \nu_{12} | \tau ) }{\eta(\tau) } \Big |^2
-\frac{2\pi}{\tau_2} (\text{Im}(\nu_{12}))^2 \, ,
}
By going to the lattice parameters, $\nu_{12} = \alpha + \tau \beta$ where $\alpha, \beta \in \mathbb{R}/\mathbb{Z}$, such that $\nu_{12} \simeq \nu_{12} + 1$ and $\nu_{12} \simeq \nu_{12} + \tau$ are identified to form the $(\alpha, \beta)$-homology cycles, it can be shown that the previous Green function is given by the lattice sum:
\ea{
G_A(\nu_{12} | \tau) = -\frac{\tau_2}{\pi} \sum_{(m,n)\neq (0,0)} \frac{e^{2\pi i (m \beta - n \alpha)}}{|m+\tau n|^2} \, ,
\label{GA}
}
which shows that this Green function is normalized such that it vanishes upon integration over the homology cycles, i.e
\ea{
\int_0^1 d\alpha \int_0^1 d\beta  \, G_A(\alpha + \tau \beta | \tau)  = 0 \, .
\label{normGA}
}

With these explicit forms for the Green function, it now becomes 
easy to calculate the $\eta^{\mu \nu}$-terms appearing in the soft expansion of the massless closed string.
At one loop the `K\"ahler form' reduces to $\mathcal{K}_1 = \frac{1}{8 \pi^2 \tau_2 |z|^2}$, and the integral to be calculated is:
\ea{
\int_{\Sigma_1} \frac{d^2 z}{4 \pi \tau_2 |z|^2} \left (1 + \sum_{\ell=1}^n \frac{\alpha'}{2}\,k_\ell \, q \, \G_1(z,z_\ell) \right ) + \Ord(q^2) \, .
} 
Using that $d^2 z = i \, dz \wedge d \bar{z} = 8 \pi^2 \tau_2 | z |^2 d\alpha d\beta$, the first term trivially gives $2\pi$. 
This is of course a more general consequence of \eqref{intK}, here shown explicitly. The heart of our problem is the second term.
To make use of \eqref{normGA} we simply make the change of variables $z= e^{2\pi i (x + \tau y)}$ and $z_\ell = e^{2\pi i (x_\ell+ \tau y_\ell)}$, whereby the second integral above becomes:
\ea{
2 \pi \sum_{\ell} \frac{\alpha'}{2} k_\ell \, q \int d x \int \, d y \,  \mathcal{G}_1 (\nu, \nu_\ell) = 0 \, .
}
The zero on the right-hand side follows immediately by choosing $\mathcal{G}_1 = G_A$ due to its translational invariance and its norm \eqref{normGA}.
However, it should also follow from any other choice, in particular from the choice $\G_1= G_B$, which was what we showed in~\cite{DiVecchia:2018dob}. But in this case it is not the integral that vanishes:
Since $G_B$ is related to $G_A$ by $G_B = G_A - \log | 2\pi \eta(\tau)^2 |^2$, i.e.~by a term independent of $z$ and $z_\ell$, the integral just yields a constant in terms of $z_\ell$. Thus the full expression vanishes, not functionally, but distributionally due to momentum conservation; i.e.~$\sum_\ell k_\ell q = - q^2 = 0$.
The lesson to learn is that by making a proper choice of coordinates on the worldsheet, the vanishing of this term becomes functionally manifest, while for other choices it is obscured due to momentum conservation. At higher genus this difference becomes much more important as translational invariance of the Green function is no longer at hand.

\section{Soft factorization at genus $h\geq2$}\label{sec:highergenus}
The higher genus generalization of \eqref{G1} takes the form~\cite{DiVecchia:2018dob}:
\ea{
\G_h(z_1,z_2)& =
\log \Big | E(z_1,z_2) \Big |^2
- \log \Big | {V_1'(0)V_2'(0)} \Big |
\label{Gh} \\
&
+ \Re \left( \int_{z_j}^{z_i} \omega _I\right) 
( 2\pi \Imt)^{-1}_{IJ} \Re \left( \int_{z_i}^{z_j} \omega _J \right)  \, ,
\nonumber
}
where $E$ is the prime form, which also has an explicit
expression in terms of the Schottky group parameters (cf.~\cite{DiVecchia:2018dob}). The relation to the usual higher genus bosonic Green function is $\G_h= G_B^{(h)}- \log \big | {V_1'(0)V_2'(0)} \big |$, which satisfies (cf.~\cite{DHoker:1988pdl}, sec. II.G.2, where $G_B^{(h)}$ is denoted $-\ln F$)
\ea{
 \partial_z \partial_{\bar{z}} G_B^{(h)}(z,w) = \pi \delta^{(2)}(z-w) - 2\pi h \,  \mathcal{K}_h(z) \, .
 }

It was shown in~\cite{DiVecchia:2018dob} that 
one can choose coordinates $V_i$ around each puncture
by using the flat metric induced from the complex plane. In the conformal gauge $ds^2 = \rho \, dz d\bar{z}$, the relation between $V_i$ and $\rho$ is given through
\ea{
|V_i'(0)| = \rho(z_i)^{-1/2} \, .
\label{coordinates}
}
On the other hand, as explained in the introduction, 
$\mathcal{K}_h$ is also a flat metric on the Riemann surface, induced from its Jacobian variety, and there is an easy way to relate it to $\rho$, namely through the Gauss-Bonnet theorem:
\ea{
\int d^2 z \sqrt{g} R(g) = - 2 \int d^2 z \,  \partial_z \partial_{\bar{z}} \log \rho = 8\pi (1-h) \, , 
}
Since we are dealing with $h\geq 2$, and since $\mathcal{K}_h$ satisfies \eqref{intK}, we get the identity
\ea{
 \partial_z \partial_{\bar{z}} \log \rho = 4\pi (h-1)  \mathcal{K}_h \, .
 \label{diffrho} 
}
This fixes $\rho$ only up to multiplication by holomorphic functions $f(z)$ and $\bar{f}(\bar{z})$ and constants $c$. It, however, uniquely fixes the Laplacian of $\G_h$ to be:
\ea{
 \partial_z \partial_{\bar{z}} \G_h(z,w) = \pi \delta^{(2)}(z-w) - 2\pi  \,  \mathcal{K}_h(z) \, .
 \label{LapGh}
 }
 This shows that the particular choice of local coordinates we here have made, gives rise to a so-called Arakelov-type Green function~\cite{DHoker:1989ima,Wentworth}.
The arbitrariness left in $\rho$, and thus in the Green function,
can be fixed by demanding a higher genus generalization of \eqref{normGA}:
Since a general solution of \eqref{diffrho} takes the form
$\rho = \tilde{\rho} \, e^{- f(z)-\bar{f}(\bar{z}) + c}$, where $\tilde{\rho}$ is a particular solution, then the general Green function has the property that
\ea{
\int d^2 z \, \mathcal{K}_h (z) \, \G_h (z, w) = 
&\int d^2 z \, \mathcal{K}_h \, \tilde{\G}_h - \int d^2 z \, \mathcal{K}_h \,\gamma(z)
\nonumber \\
&
-  \gamma(w) + c \, ,
\label{ArakelovNorm}
}
where $\gamma = (f + \bar{f})/2$ 
and $\tilde{\G}_h$ is the Green function associated with $\tilde{\rho}$.
Now, since $f$, $\bar{f}$ and $c$ are arbitrary, we can fix them such that the right-hand side above vanishes; i.e.
\ea{
\gamma (z)&:= \int d^2 w \, \mathcal{K}_h (w) \,\widetilde{\mathcal{G}}_h(z, w) \, ,
\label{f}
\\
c &:= \int d^2 z \, \mathcal{K}_h (z) \,  \int d^2 w\, \mathcal{K}_h (w) \,  \widetilde{\mathcal{G}}_h(z, w) \, .
\label{c}
}
The first identity shows that, due to \eqref{LapGh}, the Laplacian of $\gamma$ vanishes, making it a harmonic function. The Green function $\G_h$ with the choice of coordinates \eqref{coordinates}, $\rho$ given by \eqref{diffrho}, and $c$ and $\gamma$ fixed as above, \emph{is} the Arakelov Green function. It is the natural generalization of the genus one Green function $G_A$ given in \eqref{GA}.
Finally, we can show that it immediately computes the remaining integral at higher genus:
\ea{
2 \pi h \eta^{\mu \nu} \int {d^2 z}\, \mathcal{K}_h \, \prod_{\ell}{\rm e}^{ \frac{\alpha'}{2}  k_{\ell} q \G_h (z, z_{\ell})} 
= 2 \pi h \eta^{\mu \nu} + \Ord(q^2) \, ,
} 
as a consequence of the general property \eqref{intK} and of the special Arakelov property 
\ea{
\int d^2 z \,  \mathcal{K}_h \,  \mathcal{G}_h (z,w) = 0 \, ,
\label{lastintegral}
}
for any $h\geq 2$. 
Thus the entire discussion about the dilaton soft theorem made in~\cite{DiVecchia:2018dob} at one loop, directly goes through to any higher loop order, where the important role of the $\eta^{\mu \nu}$-term was also discussed.

\section{Conclusion}
We have calculated the remaining multiloop term in the soft expansion of the string amplitude involving $n$ closed string tachyons and one soft dilaton in the bosonic string and shown that it vanishes, just like at the one loop level.
This is achieved by fixing the remaining gauge freedom in~\eqref{Gh} in such a way that $\G_h$ becomes the Arakelov Green function as described in sec.~\ref{sec:highergenus}. 
It implies that scattering amplitudes involving one soft dilaton with momentum $q$ factorizes as follows~\cite{DiVecchia:2018dob} (see~\cite{Green:2019rhz} for a recent leading-order application):
\begin{widetext}
\ea{
\mathcal{M}_{n;\phi} (k_i; q) = 
\frac{\kappa_d}{ \sqrt{d-2} } 
\left[ - \sum_{i=1}^n \frac{m_i^2}{k_i q} {\rm e}^{q \partial_{k_i}}  
+ \frac{d-2}{2} g_d \frac{\partial}{\partial g_d} - \sqrt{\alpha'} \frac{\partial}{\partial \sqrt{\alpha'}}   
+ q_\mu \sum_{i=1}^n K_i^\mu
\right] \mathcal{M}_n(k_i) + \Ord(q^2) \, ,
}
\end{widetext}
where $\mathcal{M}_{n;\phi}$ and $\mathcal{M}_{n}$ are the all-loop amplitudes with and without the soft dilaton, $\kappa_d$ and $g_d$ are, respectively, the $d$-dimensional gravitational and string coupling constants, $m_i$ are the masses of the hard external states carrying momentum $k_i$, $\alpha'$ is the inverse string tension, and $K_i^\mu$ is the momentum space generator of special conformal transformations, all explicitly defined in~\cite{DiVecchia:2018dob}. We have compactified $26-d$ spacetime dimensions, but the choice of compactification geometry is irrelevant to our final result. The role of IR divergences was discussed in~\cite{DiVecchia:2018dob}, and shown to be irrelevant for this factorization property of the amplitude in $d$-dimensions greater than four.


\appendix
\section{
From Schottky to Jacobi at one loop}
\label{SJ}

From the explicit formulas in Sec.~B of~\cite{DiVecchia:2018dob}, 
we readily derive the explicit expressions for the abelian differentials, period `matrix', and prime form at one-loop in the Schottky parametrization:
\ea{
\omega (z) &= \frac{1}{z} \, , \qquad \qquad \ 
\bar{\omega} (\bar{z}) = \frac{1}{\bar{z}} \, ,
\\
2\pi i \tau &= \log \kappa \, , \qquad \qquad  \left ( \kappa = e^{2\pi i \tau} \right )
\\
E(z_1,z_2) &= e^{i \pi (\nu_1 + \nu_2)} \,  2 i \, \sin(\pi \nu_{12}) 
\nonumber \\
& \ \times \prod_{n=1}^\infty \frac{(1-\kappa^n e^{2\pi i \nu_{12}})(1-\kappa^n e^{- 2\pi i \nu_{12}})}{(1-\kappa^n)^{2} } \, , 
}
where $\kappa$ is the genus one Schottky group multipler, and $\nu_{ij}=\nu_1 - \nu_j$ with $z_i = e^{2\pi i \nu_i}$. We note that at genus one, the `period matrix' is just the usual modular parameter, and the prime form has, apart from the prefactor $e^{i \pi (\nu_1 + \nu_2)} = \sqrt{z_1 z_2} $, translational invariance in the $\nu_i$ variables, i.e. 
\ea{
E(z_1,z_2) =  i \, \sqrt{z_1 z_2} \,  \mathcal{E}(\nu_1-\nu_2 | \tau) \, .
}
$\mathcal{E}$ can be written in terms of the Jacobi theta function and its derivative $\partial_\nu \theta_1(\nu |\tau) := \theta_1'(\nu|\tau)$ as follows
\ea{
\mathcal{E}(\nu | \tau) = 2\pi\, \frac{ \theta_1( \nu | \tau ) }{\theta_1'(0 |  \tau) } \, ,
}
as easily seen from the product representation of  $\theta_1$:
\ea{
\theta_1 (\nu | \tau) &= 2 \kappa^{1/12} \sin (\pi \nu) \eta(\tau) \prod_{n=1}^\infty(1 - 2 \kappa^n \cos (2\pi \nu) + \kappa^{2n}) \, ,
\nonumber\\
\theta_1'(0 |  \tau) &= 2 \pi \eta(\tau)^3 \, ,
\label{thetaprime}
}
where we introduced the Dedekind eta function, 
\ea{
\eta(\tau) = \kappa^{1/24}\prod_{n=1}^\infty (1-\kappa^n) \, .
\label{Dedekind}
}
Inserting these expressions in the $h$-loop expression for the Green function one readily finds Eq.~\eqref{G1}.


\end{document}